\documentclass[twocolumn,aps,pra,showpacs,superscriptaddress,
longbibliography,10pt]{revtex4-1}

\usepackage{bm}
\usepackage{amssymb}
\usepackage[usenames,dvipsnames]{color}
\usepackage{graphicx}
\usepackage{dcolumn}
\usepackage{hyperref}
\usepackage{natbib}
\usepackage[caption=false]{subfig}
\usepackage{float}
\usepackage{times}
\usepackage{booktabs}		
\usepackage{mathtools}
\usepackage{amsmath,array}
\usepackage{relsize}
\usepackage{enumerate}
\usepackage{color}

\usepackage{listings}
\usepackage{epstopdf}
\usepackage{braket}

\usepackage[utf8]{inputenc}

\newcommand{\ncyc}{n_{\mathrm{cyc}}}

\newcommand{\Up}{U_\mathrm{p}}

\DeclareMathAlphabet{\bi}{OML}{cmm}{b}{it}

\newcommand{\diff}{\mathrm{d}}

\newcommand{\pabl}[2]{\frac{\partial #1}{\partial #2}}

\newcommand{\Ip}{I_{\mathrm{p}}}

\newcommand{\beq}{\begin{equation}}
\newcommand{\eeq}{\end{equation}}

\newcommand{\vxc}{v_\mathrm{xc}}

\newcommand{\vKS}{{v}_\mathrm{KS}}
\newcommand{\vKStilde}{{\tilde v}_\mathrm{KS}}

\newcommand{\Wcmcm}{W/cm$^2$}

\begin{document}

\title{High-order harmonic generation in solid slabs beyond the single active electron}
\author{Kenneth K. Hansen}
\affiliation{Department of Physics and Astronomy, Aarhus University, DK-8000, Denmark}
\author{Tobias Deffge}
\author{Dieter Bauer}
\affiliation{Institute of Physics, University of Rostock, 18051 Rostock, Germany}

\begin{abstract}
High-harmonic generation by a laser-driven solid slab is simulated using time-dependent density functional theory. Multiple harmonic plateaus up to very high harmonic orders are observed already at surprisingly low field strengths. The full all-electron harmonic spectra are, in general,  very different from those of any individual Kohn-Sham orbital. Freezing the Kohn-Sham potential instead is found to be a good approximation for the laser intensities and harmonic orders considered. The origins of the plateau cutoffs are explained in terms of band gaps that can be reached by Kohn-Sham electrons and holes moving through the band structure. 
\end{abstract}
\date{\today}
\maketitle

\section{Introduction}\label{intro}
High-harmonic generation (HHG) in the gas phase laid the foundation of attosecond science, opening up the possibility to study ultrafast processes on the sub-femtosecond scale directly in the time domain \cite{attoKrausz,attoCalegari}. The basic ingredient in the description of HHG in rarefied gases is the single-atom polarization, which then might be used as a source in Maxwell's equations to calculate the propagation of the harmonic radiation through the gas \cite{PropagationPriori,propagationGaarde}. The overall shape of an HHG spectrum generated by atoms is mainly governed by the ionization potential $\Ip$ and the laser pulse form, in particular its electric field amplitude $F_0$ and the laser frequency $\omega_0$, which determine the ponderomotive energy $\Up=F_0^2/4\omega_0^2$ (atomic units where electron mass $m_e$, charge $|e|$, and $4\pi\epsilon_0$ are unity are used, unless indicated otherwise). In particular, the celebrated cutoff for the HHG plateau $\omega_{\max}=\Ip + 3.17\Up$ is determined by only these few parameters and can be understood in terms of the ``three-step model'' \cite{LewensteinPhysRevA.49.2117,RecollCorkumPhysRevLett.71.1994} where the electron is (1) assumed to be released due to tunneling, (2) oscillates in the laser field, and (3) swings back to the parent ion where it recombines upon emitting its kinetic energy plus the ionization potential as a single harmonic photon.  

Recently, strong-field physics in solids, in particular HHG, got into the focus of attention \cite{Ghimire2011,GhimirePhysRevA.85.043836,Ghimire0953-4075-47-20-204030,HutterReviewLPOR:LPOR201700049}, with potential applications in light-driven electronics \cite{Schiffrin2013,Schultze2013,Garg2016,Higuchi2017,Schultze1348}, efficient and compact terahertz radiation sources \cite{SchubertO.2014}, polarization and phase shaping of the emitted radiation \cite{LangerF.2017}, ultrafast dielectric optical switching \cite{Sommer2016}, and all-optical measurements of band structure \cite{VampaPhysRevLett.115.193603} and dynamics therein \cite{Luu2015,Hohenleutner2015,Hassan2016,Lucchini916,Ndabashimiye2016,You2017}. The combination of periodic drivers with spatially periodic systems opens up the new field of ``Floquet matter'' and its topological properties \cite{FaisalPhysRevA.56.748,OkaPhysRevB.79.081406,BucciPhysRevB.96.041126,DimiPhysRevA.95.063420,MoisPhysRevA.91.053811}.   Compared to strong-field physics in atoms, the situation is richer in condensed matter because the trivial (and for all atoms equal) dispersion relation of a free electron, $E(k)=k^2/2$, is replaced by a target-dependent band structure so that the relations between (crystal) momentum, electron velocity and hole velocity are not as simple as in the atomic case.  Clearly, the laser intensity must remain below the damage threshold to employ the band structure for HHG in solids (unless one is interested in plasma-based harmonics from the surface \cite{Tsak1367-2630-8-1-019,BehmkePhysRevLett.106.185002}). Nevertheless it turned out that many of the semi-classical concepts that have been developed for strong-field physics in the gas phase can be transferred to the interaction of laser radiation  of much lower laser intensities with semi-conductors or insulators as long as the photon energies are much smaller than the band gap between valence and conduction band. Further, the band structure might be such that effective masses are small or dispersion relations are relativistic-like so that laser fields that can be considered weak by strong-field standards in vacuum are effectively strong in a solid.    

The attempts to understand HHG in solids necessarily require concepts from two previously rather distinct disciplines: strong-field physics and condensed matter physics. Theory papers on the subject reflect this method-wise: HHG in solids has been investigated in the strong-field-way employing a three-step-like approach \cite{VampaTutorial0953-4075-50-8-083001} or solving the time-dependent Schr\"odinger equation for single electrons in a periodic potential and a laser field \cite{Numico0953-4075-33-13-319,gaarde_HHG,WuPhysRevA.94.063403,SatoPhysRevB.89.224305,YuPhysRevA.94.013846,Hawkins2015,ishikawa_HHG,chinese_hhg_TDPI}, 
or in the condensed-matter way  using semi-conductor Bloch equations \cite{Semiclassical_many_elec,WismerPhysRevLett.116.197401,HutterReviewLPOR:LPOR201700049}. Time-dependent density functional theory (TDDFT) \cite{UllrichBook,QSFQDBook}  lies somewhat at the boundary as it has been used in both communities for a long time \cite{Otobe0953-8984-21-6-064224,TancPhysRevLett.118.087403,OtobePhysRevB.94.235152,Tancogne-Dejean2017}.

Concerning HHG in solids, some obvious questions arise: (i) How do HHG spectra from solids look like, and which cutoffs are observed? (ii) Which electrons do the HHG? (iii) Is electron-electron interaction important? (iv) Are there surface effects in HHG? (v) How does the coupling to phonons influence HHG spectra? (vi) How does the incident pulse and the harmonic radiation propagate through the solid? In this paper, we will address points (i)--(iii) whereas items (iv)--(vi) are postponed to forthcoming work.   

In order to pinpoint the essentials of HHG by linearly polarized laser pulses in solids we employ a TDDFT model for a linear chain of $N$ ions in a laser field. In that way we go beyond single-active electron models and take electron-electron interaction into account, at least on a mean-field level. The TDDFT model is introduced in Sec.~\ref{model}. We show the band structure for the particular parameters used for the subsequent simulations with the laser field in Sec.~\ref{bandstruct}, and discuss the Bloch oscillations in a static electric field in Sec.~\ref{Blochoscillations}. HHG spectra are presented in Sec.~\ref{results}, in particular their dependence on the number of ions $N$ in Sec.~\ref{numbofions}, the difference between dynamic and frozen Kohn-Sham (KS) potential in Sec.~\ref{dynamicandfrozen}. Plateaus and cutoffs are discussed in Sec.~\ref{discussion}. We summarize in Sec.~\ref{sum}.

\section{(TD)DFT model}\label{model}
We consider a linear chain of $N$ ions of charge $Z$ at positions $x_i$, separated by the lattice constant $a$, 
\begin{align}
x_i  &=  \left[i- \frac{1}{2}(N-1)\right]a, \label{eq:equidistant}
\end{align}
generating the attractive potential for the electrons 
\begin{align}
v_\mathrm{ion}(x) &=  - \sum_{i=0}^{N-1} \frac{Z}{\sqrt{(x-x_i)^2 + \epsilon}} \label{eq:ionpot} .
\end{align}
The smoothing parameter $\epsilon$ is introduced to soften the 3D Coulomb singularity in a 1D treatment, see, e.g., \cite{Eberly}. 
The ionic potential \eqref{eq:ionpot} enters the KS potential
\begin{align} 
\vKS[\{n_{\sigma}\}](x) &= v_\mathrm{ion}(x) +  u[n](x) + \vxc[\{n_{\sigma}\}](x)
\end{align}
in the KS equation for the KS orbitals $ \varphi_{\sigma,i}(x)$,
\begin{align} 
\epsilon_{\sigma,i} \varphi_{\sigma,i}(x) & =  \left( - \frac{1}{2} \pabl{^2}{x^2} +  \vKS[\{n_{\sigma}\}](x) \right)\varphi_{\sigma,i}(x), \label{eq:tiKS}
\end{align}
together with the Hartree potential
\begin{align} 
u[n](x) &= \int \frac{n(x')\, \diff x'}{\sqrt{(x-x')^2 + \epsilon}} \label{eq:hartree}
\end{align}
and the exchange-correlation potential in local spin-density (LSD) approximation 
\begin{align}
\vxc[\{n_{\sigma}\}](x) & \simeq  v_\mathsf{x}[\{n_{\sigma}\}](x) = -\left( \frac{6}{\pi} n_\sigma(x)\right)^{1/3}. \label{eq:lsd}
\end{align}
The spin densities and the total density read
\begin{align}
n_\sigma(x) &= \sum_{i=0}^{N_{\sigma}-1} |\varphi_{\sigma,i}(x)|^2, \quad n(x)=\sum_{\sigma=\uparrow,\downarrow} 
n_\sigma(x),
\end{align}
respectively, with $N_\sigma$ the number of electrons of spin $\sigma$, and $N_e=N_\uparrow+N_\downarrow$ the total number of electrons. We use the LSD exchange expression for the 3D electron gas because in what follows we want to mimic 3D electrons that are driven in the polarization direction of a linearly polarized laser rather than a true 1D electron system. The LSD correlation part is neglected, as it is not expected to affect the qualitative features in our results. In this work, we restrict ourselves to even $N$ for which the system is spin-neutral and the local density approximation (LDA) would actually suffice. However, for odd $N$ this is not the case anymore, and we want to keep the theory and our code flexible right from the beginning. Further, we are not using periodic boundary conditions in order to monitor the escape of electrons from the solid as a whole, and other surface effects.

\subsection{Two exemplary band structures} \label{bandstruct}
\begin{figure}
\centering
\includegraphics[width=\columnwidth]{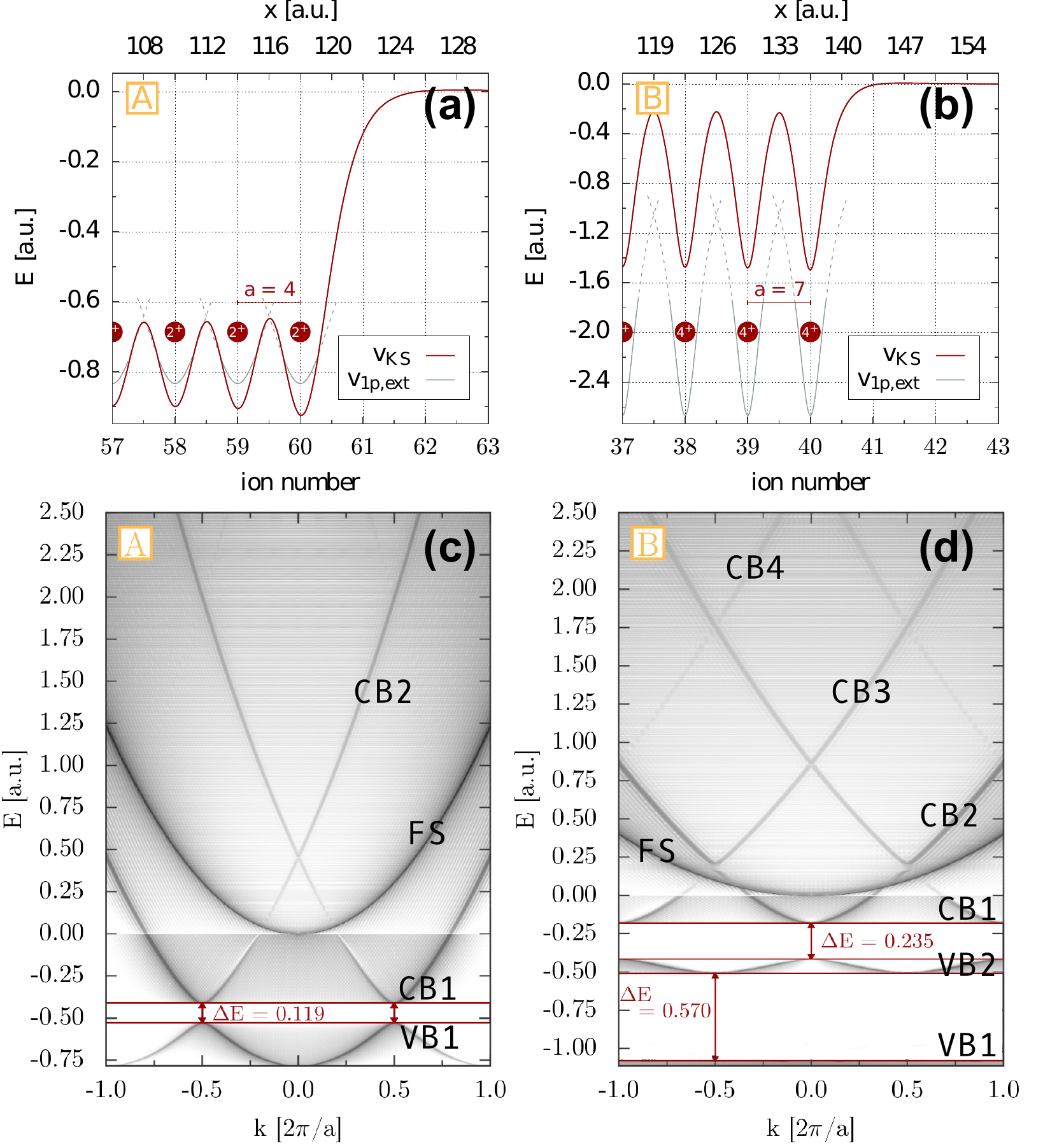}
\caption{Right edges of the KS potentials for (a) $N=60$, $Z=2$, $a=4$, $\epsilon=5.75$ (system {A}) and (b) $N=40$, $Z=4$, $a=7$, $\epsilon=2.25$ (system {B}). The single-ion potentials $v_\mathrm{1p,ext}=-Z/\sqrt{(x-x_i)^2 + \epsilon}$ are indicated by black dashed and dotted lines. The corresponding band structures in (c) and (d) are calculated from the spatial Fourier transforms of the occupied and unoccupied KS orbitals, plotted at the level of their KS orbital energy. Occupied bands are labeled VB$i$, $i=1,2$, initially empty bands CB$i$, $i=1,2,3,4$. Because of the finiteness of the slab there is also the free-space (FS) dispersion $k^2/2$ visible in both (c) and (d).}
\label{fig:sysAandB}
\end{figure}

The four adjustable parameters in our model are the number of ions $N$, the lattice constant $a$, the smoothing parameter $\epsilon$, and the ion charge $Z$. They can be tuned to obtain the desired band fillings or band gaps, for instance. Figure~\ref{fig:sysAandB} shows examples for $N=60$, $Z=2$, $a=4$, $\epsilon=5.75$ (system {A}), and  $N=40$, $Z=4$, $a=7$, $\epsilon=2.25$ (system {B}). In order to render the solid neutral there are $ZN=N_e$ electrons and, because of spin-neutrality, $N_\uparrow=N_\downarrow=N_e/2$ in both systems. The KS equation \eqref{eq:tiKS} was solved on an equidistant spatial grid of spacing $\Delta x=0.2$ by propagating KS orbitals according to the time-dependent KS equation \eqref{eq:tdKS} below in imaginary time and orthonormalizing each timestep \cite{QSFQDBook}.
In Fig.~\ref{fig:sysAandB}a,b, the right edges of the KS potentials for the two systems are shown. The main difference is the different height of the potential in between the ions as compared to the continuum threshold $E=0$. Panels (c) and (d) show the corresponding band structures. With $Z=2$ electrons per ion in system {A}, the lowest band is fully populated (valence band, VB1), and all higher ones are empty (conduction bands CB1, CB2). Because we consider a finite solid slab there is also the free-space dispersion parabola $k^2/2$ present (FS). The $Z=4$ electrons per ion in system {B} yield two completely filled bands VB1, VB2 and empty conduction bands CB1--CB4, besides the free-space parabola. We expect the highest occupied band and the lowest conduction band to determine most of the relevant strong-field dynamics of solids, the smallest band gap between them playing the role of what would be the ionization potential in the gas-phase case \cite{VampaTutorial0953-4075-50-8-083001}.  By changing the parameters $N,a,\epsilon,Z$, the band structure can be tuned to the desired shape. In particular, the band gap can be adjusted to values of interest. Due to the different fillings the minimum band gap can be at the Brillouin-zone boundary $k=\pi/a$ (system {A}) or $k=0$ (system {B}). For $Z=1$ (or $Z=3$) electron(s) per ion half-filled bands and a metallic behavior would be obtained. In the following, we will restrict ourselves to present results obtained for system {B}, which yields an effective KS potential similar to the one used in  \cite{gaarde_HHG}.

The interaction of our model with external (laser) fields is simulated using TDDFT. In the adiabatic approximation, the stationary KS equation \eqref{eq:tiKS} is replaced by the time-dependent KS equation
 \begin{align} 
i\pabl{}{t} \varphi_{\sigma,i}(x,t) & =  \Bigl( - \frac{1}{2} \pabl{^2}{x^2}- i A(t) \pabl{}{x} \nonumber \\
&\quad +  \vKStilde[\{n_{\sigma}\}](x,t) \Bigr)\varphi_{\sigma,i}(x,t), \label{eq:tdKS}
\end{align} 
where
\begin{align} 
\vKStilde[\{n_{\sigma}\}](x,t) &= v_\mathrm{ion}(x) +  u[n](x,t) + \vxc[\{n_{\sigma}\}](x,t) 
\end{align}
with $A(t)$ the vector potential of the laser field in dipole approximation, and the time-dependent density used in the expressions for the Hartree potential \eqref{eq:hartree} and the exchange  potential \eqref{eq:lsd}. The KS orbitals were propagated in time according to the time-dependent KS equation \eqref{eq:tdKS} using  the Crank-Nicolson method (with a predictor-corrector step) \cite{QSFQDBook}.  The system starts at time $t=0$ from the ground state with spin densities $n_\sigma(x,0)=n_{\sigma 0}(x)$ and total density $n(x,0)=n_0(x)$.

\subsection{Bloch oscillations} \label{Blochoscillations}
\begin{figure}
\centering
\includegraphics[width=\columnwidth]{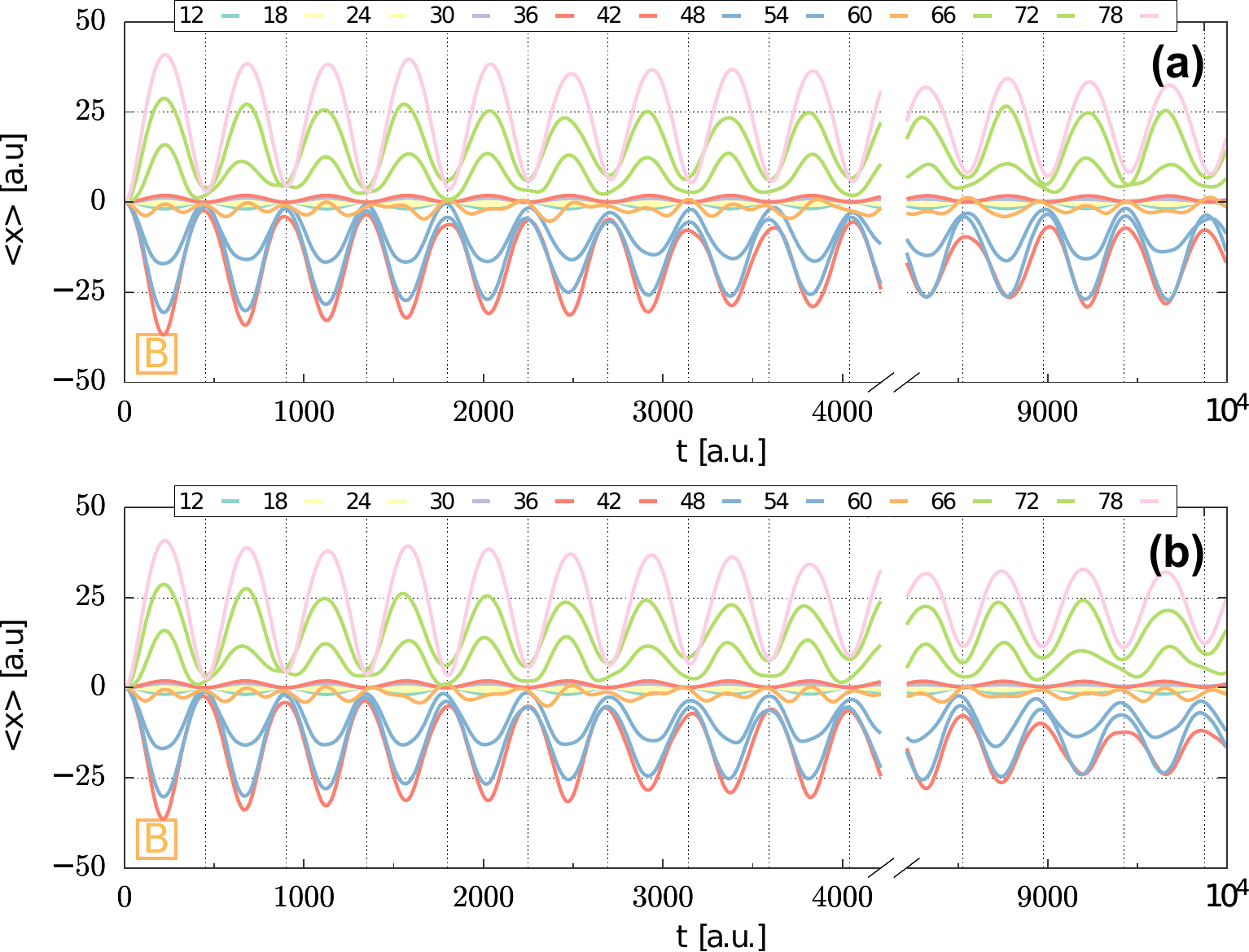}
\caption{Bloch oscillations in system B for a small, instantaneously switched-on electric field $F=0.002$.  Position expectation values \eqref{eq:posexpect} for exemplary KS orbitals are shown from the full TDDFT simulation (a) and the frozen-KS-potential simulation (b). Thin vertical lines indicate multiples of the expected Bloch period $2\pi/\Omega_\mathrm{B} \simeq 449$. }
\label{fig:bloch}
\end{figure}

It is instructive to calculate the response of the KS system to a small, static electric field $F=-\partial_t A(t)$, i.e., $A(t)=-Ft$. A single electron in a periodic potential in the presence of a static electric field will undergo Bloch oscillations of frequency $\Omega_\mathrm{B}=aF$. However, in a completely filled valence band electrons at the top edge of the band will oscillate oppositely to electrons at the bottom because of the opposite velocities $v(k)=\partial_k E(k)$. This is seen   in Fig.~\ref{fig:bloch}, where the position expectation values 
\begin{align}
\langle x_{\sigma, i}\rangle (t) = \int\diff x\, x \left|\varphi_{\sigma, i}(x,t)\right|^2 \label{eq:posexpect}
\end{align} 
for, e.g., spin up (spin-down is the same) are plotted for representative KS orbitals {\em vs} time for the case of an instantaneously switched-on electric field $F=0.002$. The upper panel (a) shows the full TDDFT result with the KS potential updated each time step according to the instantaneous electron density, panel (b) the one for a frozen KS potential
\begin{align} 
{\vKS}_0(x) &= v_\mathrm{ion}(x) +  u[n_0](x) + \vxc[\{n_{\sigma 0}\}](x).
\end{align}
Freezing the KS potential is equivalent to a non-interacting electron simulation where all electrons move independently in a given, static, effective potential (for whose calculation electron-electron interaction has been taken into account though). The Pauli principle is fulfilled for both dynamic and frozen KS potential:  if the KS orbitals are orthogonal at $t=0$ (which they are by construction) they will stay orthogonal during time-propagation as long as the KS Hamiltonian is invariant under the exchange of (like spin) KS particles.

For small enough field strengths $F$ there is almost no difference between the full and the frozen-KS-potential result. KS orbitals 0--39 belong to the lowest, completely filled band VB1 of system B. Their oscillation amplitudes are small compared to the second completely filled band VB2 because of the higher effective mass $m^*$ of band VB1. KS orbitals 40--59 of VB2 show excursions in $-F$ direction because of their dispersion relation approximately $\propto k^2$ (positive $m^*$), orbitals 61--79 oscillate in opposite direction (opposite band curvature, negative $m^*$). In this way the KS orbitals remain orthogonal, and the Pauli principle can be fulfilled. The sum of all the individual KS dipoles almost cancels both in the full-KS-potential and the frozen-KS-potential result. Hence, our model reproduces the experimental fact that Bloch oscillations are usually (i.e., without purposefully designed superlattices \cite{Leo0268-1242-13-3-003}) not seen experimentally, however, not because of scattering or dissipation processes that are faster than the Bloch period  but even for non-interacting electrons in filled bands because of destructive interference of the individual KS dipoles. Such interferences are expected to play a role in HHG in solids as well. In fact, as discussed below in  Sec.~\ref{discussion}, the HHG spectrum produced by an individual KS orbital (e.g., the highest occupied one) is in general  very different from the measurable spectrum, generated by all electrons.

\section{HHG spectra}\label{results}
We calculate HHG spectra either from the modulus square of the Fourier-transformed position expectation value (or dipole) $\langle x \rangle(\omega)=\mathrm{FFT}[\langle x\rangle (t)]$ where $\langle x\rangle (t)=\sum_{\sigma,i} \langle x_{\sigma, i}\rangle (t)$ with $\langle x_{\sigma, i}\rangle (t)$ according \eqref{eq:posexpect}, or from the Fourier-transformed current $j(\omega)=\mathrm{FFT}[j(t)]$ where $j(t)=\sum_{\sigma,i} \int\diff x\, j_{\sigma, i}(x,t)$ with $ j_{\sigma, i}(x,t) = -i[\varphi^*_{\sigma,i}(x,t)\partial_x \varphi_{\sigma,i}(x,t) - \varphi_{\sigma,i}(x,t)\partial_x \varphi^*_{\sigma,i}(x,t)]/2$.

\subsection{How many ions make bulk?}\label{numbofions}
Keeping the lattice constant $a=7$, the ion charge $Z=4$, and the smoothing parameter $\epsilon=2.25$ of system B above but varying the number of ions $N$, we can study how many ions are needed in laser polarization direction of the laser to yield the ``converged'' bulk HHG spectrum. We assume a laser field with the vector potential
\begin{align}
A(t) &= A_0 \sin^2\left(\frac{\omega t}{2 \ncyc} \right) \sin\omega_0 t   \label{eq:vpot}
\end{align}
for $0<t<\ncyc 2\pi/\omega_0$ and zero otherwise.

\begin{figure}
\centering
\includegraphics[width=\columnwidth]{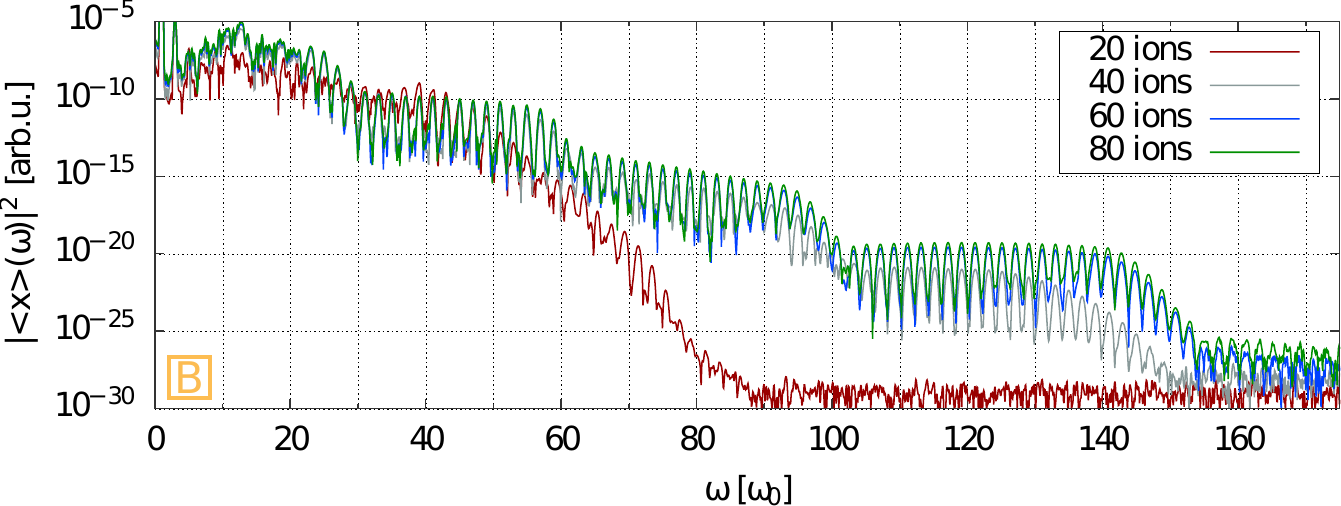}
\caption{HHG spectra for different numbers of ions $N$ in the linear chain, frozen KS potential, and an $\ncyc= 15$-cycle $\sin^2$ pulse with $\omega_0 = 0.023$ (corresponding to $\lambda \simeq 2\,\mu$m) and $A_0 = 0.24$ (corresponding to a laser intensity $\simeq 10^{12}$\,\Wcmcm). }
\label{fig:size_difference}
\end{figure}

The number of lattice sites in the system clearly affects the HHG signal produced. 
In Fig.~\ref{fig:size_difference}, HHG spectra are presented for $N=20,40,60,80$ ions.
For instance, $N= 40$ ions generate an HHG spectrum where all the structures of the well converged bulk result $N=80$ are present up to harmonic order $ \simeq 130$ for which the yield is already many orders of magnitude below the fundamental. As discussed below, such high harmonics involve electrons moving in  high-lying conduction bands where their excursions in position space are too large to be supported by smaller crystal sizes.

\subsection{HHG with dynamic and frozen KS potential} \label{dynamicandfrozen}
Similar to Sec.~\ref{Blochoscillations} above for the Bloch oscillations, we do not expect a large difference between results obtained with full, dynamic TDDFT, i.e., with the KS potential updated each time step, and those obtained with a frozen, ground state KS potential as long as the system stays close to its ground state configuration (which should be the case well below the damage threshold of the solid).

\begin{figure}
\centering
\includegraphics[width=\columnwidth]{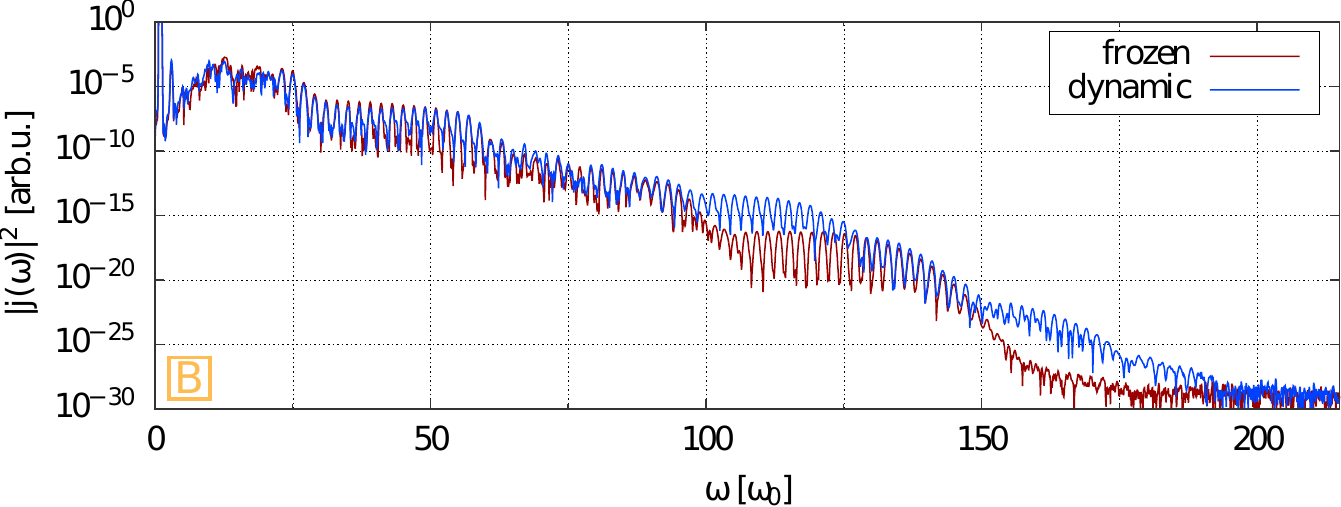}
\caption{HHG spectra for system B (with $N=40$) for frozen and dynamic KS potential, calculated from the Fourier-transformed total current. The laser parameters are the same as in Fig.~\ref{fig:size_difference}.}
\label{fig:frozenvdynamic}
\end{figure}

Figure~\ref{fig:frozenvdynamic} shows HHG spectra for the parameters of Fig.~\ref{fig:size_difference} and $N=40$ ions for frozen and dynamic KS potential. Differences between the two spectra are only visible from harmonic order $\simeq 100$ on. Actually, it is not clear which of the two spectra is closer to the unknown exact result. Benchmark results from a solution of the $N_e=160$-electron time-dependent Schr\"odinger equation are impossible to obtain, and converged results from other methods such as time-dependent multi-configurational Hartree-Fock are still too costly \cite{zangh0953-4075-37-4-004,QSFQDBook}.  The time-dependent KS equation, as a non-linear partial differential equation where the KS orbitals feed into the effective potential, may generate artificial high-order harmonics, as observed, e.g., for HHG in He \cite{tdrnot4}. The unknown, exact xc potential would care about removing the artificial harmonics and placing the physical ones like those due to single-photon, nonsequential double recombination \cite{KovalPhysRevLett.98.043904,HansenPhysRevA.96.013401} or simultaneous HHG with different charge states, for instance. Anyhow, the discrepancies between full and frozen-KS-potential result in  Fig.~\ref{fig:frozenvdynamic} are at harmonic orders where the yield is already very low. Hence our results up to practically relevant harmonic orders are ``robust'' in the sense that the fine details of how accurate electron-electron interaction is taken into account via the chosen xc potential are of minor importance. 

\subsection{Discussion of the HHG spectra} \label{discussion}
Semi-classical methods have been proposed to predict the cutoffs observed in HHG spectra from solids \cite{VampaTutorial0953-4075-50-8-083001,Semiclassical_many_elec,ishikawa_HHG}. These methods are based on the motion of a state that is initially localized in momentum space at (or around) the crystal momentum $k_0$ and then following the external field according to the adiabatic theorem, i.e.,  $k(t) = k_0 + A(t)$. This means that the state moves  along the band with the vector potential, which produces already so-called intraband harmonics due to anharmonicities, i.e., because realistic bands are not perfectly parabolic \cite{Luu2015}. Transitions to higher bands are most probable when the band gap is smallest. Hence, one might assume in the semi-classical modeling that such transitions occur whenever the minimum band gaps between successive bands are reached. After the transition happened, the electron moves in its new band but may recombine with the hole it left in its initial band. This scenario is an analogue of the three-step model for HHG in the gas phase \cite{VampaTutorial0953-4075-50-8-083001}. A maximum so-called interband HHG energy is then determined by the maximum energy difference between the conduction band the electron reached and the valence band where it started, at a specific $k$ value that is accessible from at least one initial $k_0$.

\begin{figure}
\centering
\includegraphics[width=\columnwidth]{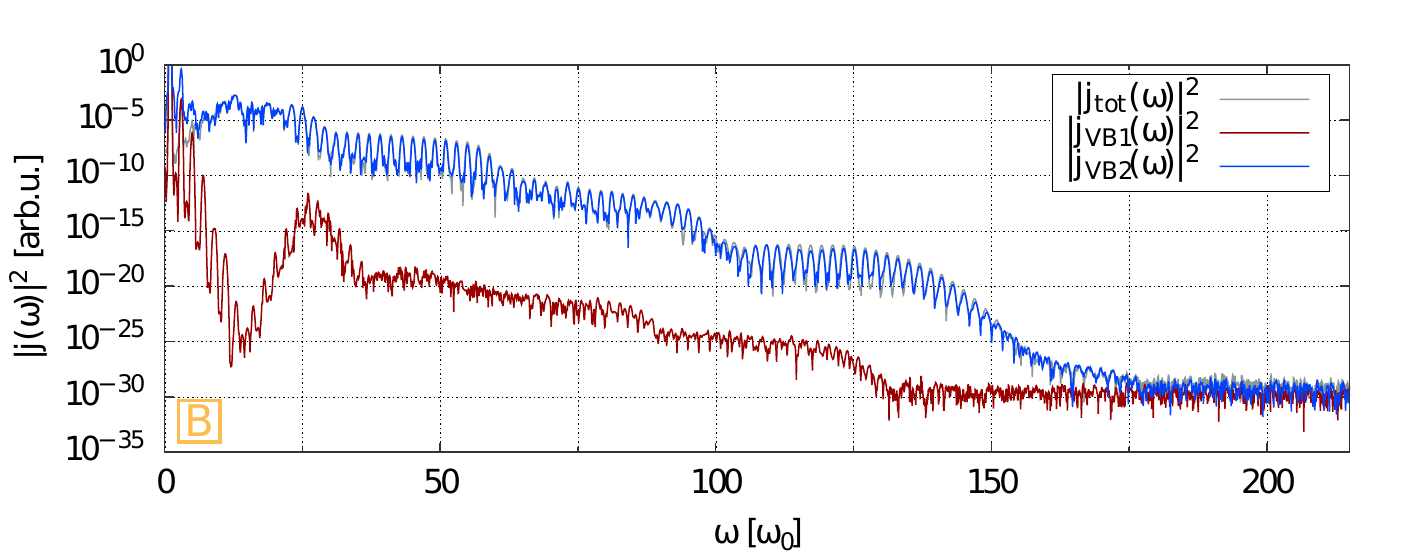}
\caption{HHG spectra for system B calculated from the total current, the current from all electrons in VB1 and from VB2. The laser parameters are the same as in Figs.~\ref{fig:frozenvdynamic} and \ref{fig:size_difference}. The KS potential was frozen.}
\label{fig:spectra_bands}
\end{figure}

System B has initially two completely filled bands VB1 and VB2, see Fig.~\ref{fig:sysAandB}d.  When determining which states contribute to the total current and thus to HHG, we first combine the current originating from VB1 and VB2 separately. The corresponding spectra are presented in Fig.~\ref{fig:spectra_bands}, together with the full spectrum from the total current.
It is seen that the contribution from KS orbitals that were initially in the lowest band is insignificant for most frequencies, except for the intraband harmonics $< 10 \omega_0$. This is expected because of the large band gap suppressing electrons in VB1 from moving up in the band structure, therefore not enabling them to create interband high harmonics there.
The total spectrum coincides almost perfectly with the spectrum from the KS orbitals initially in VB2 alone.

\begin{figure}
\centering
\includegraphics[width=0.9\columnwidth]{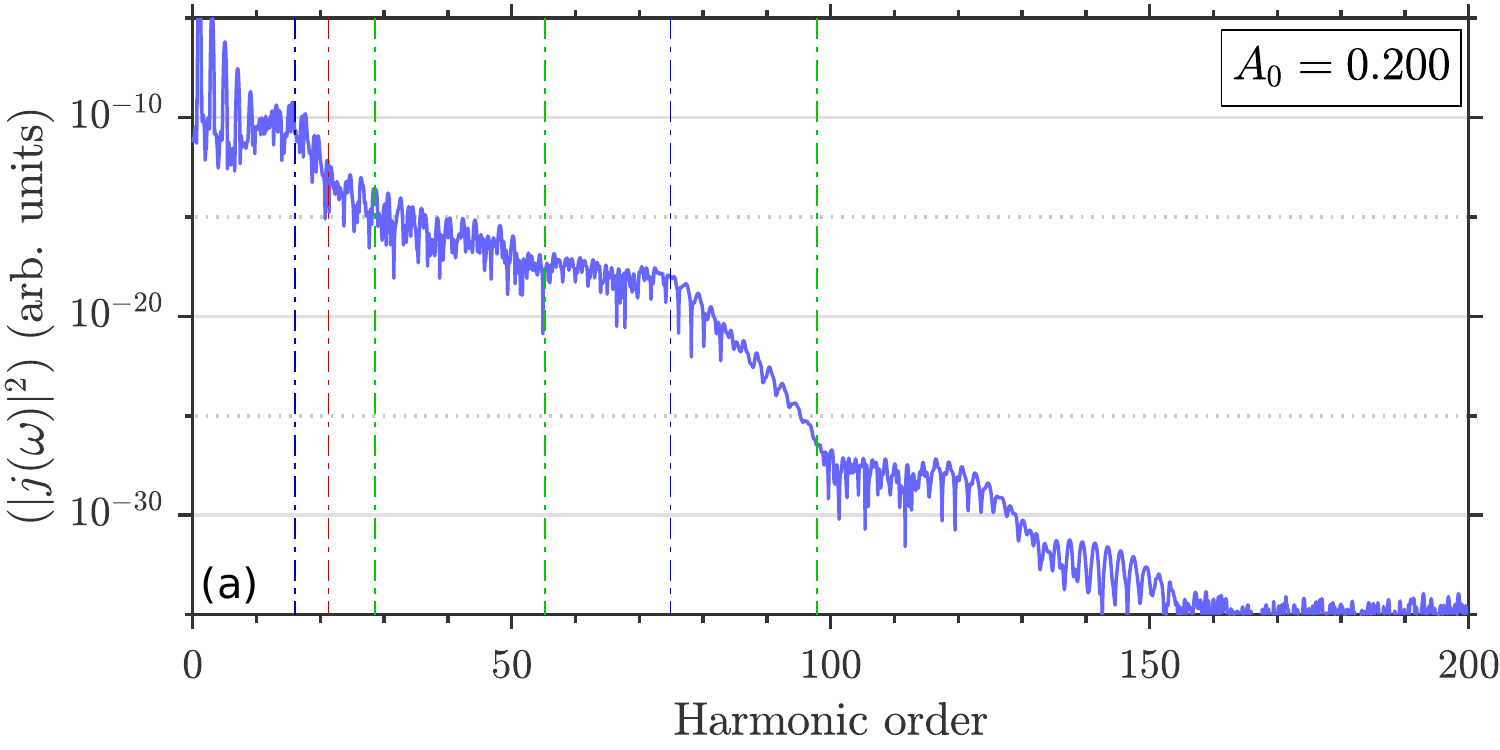}
\includegraphics[width=0.9\columnwidth]{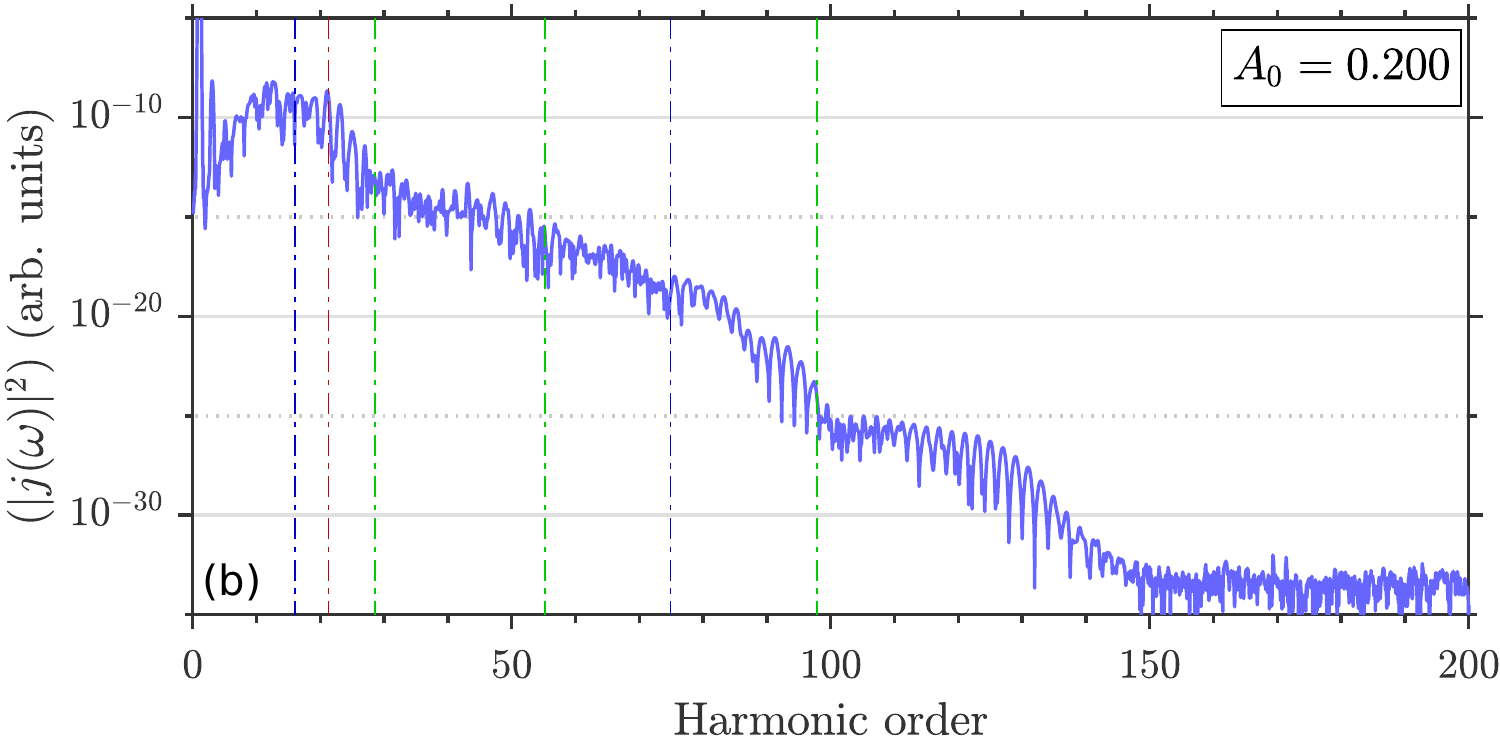}
\includegraphics[width=0.9\columnwidth]{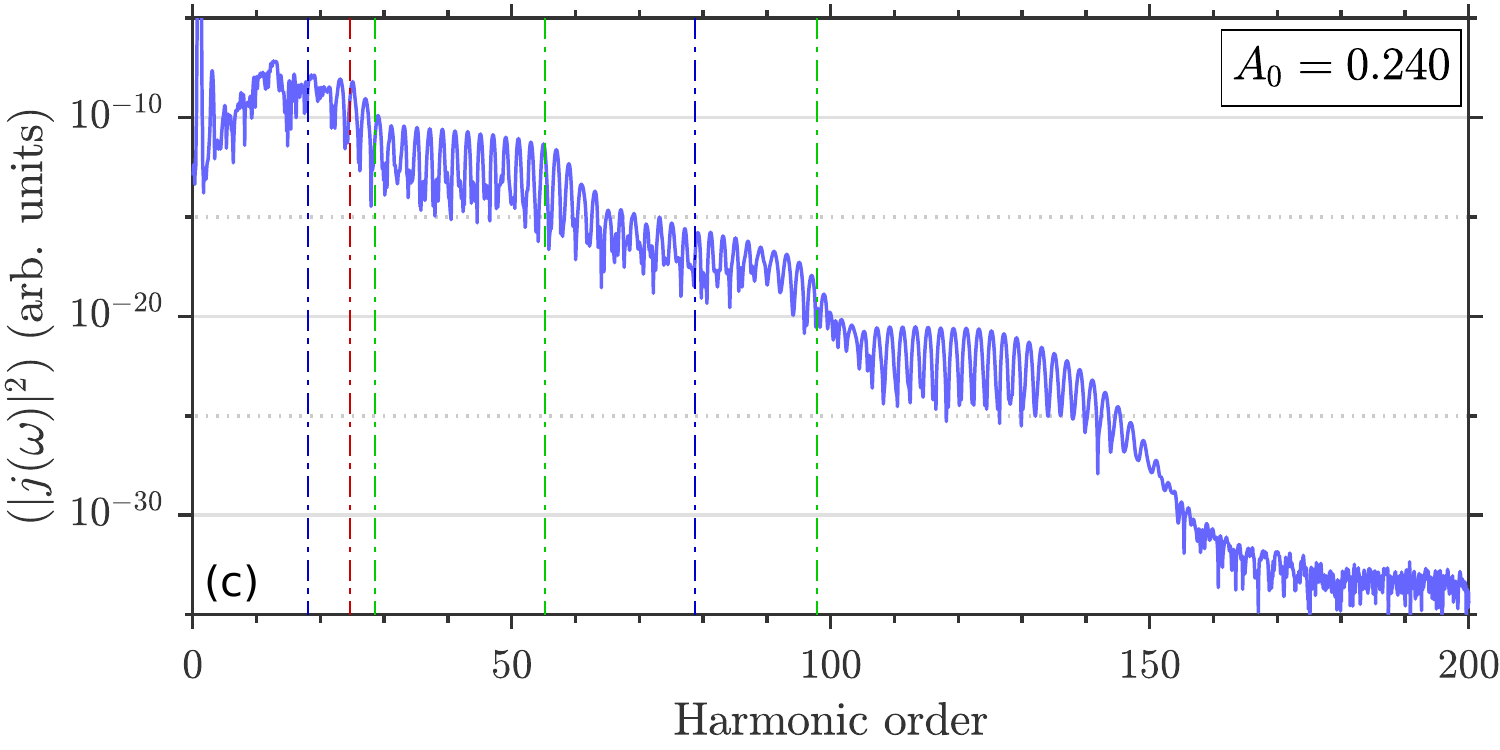}
\caption{HHG spectra from system B obtained for the same pulse shape as in Fig.~\ref{fig:spectra_bands} but different  vector potential amplitudes. Panel (a) shows the HHG spectrum from the highest occupied orbital.  The first dashed blue, vertical line from the left indicates the energy difference between the highest valence band VB2 and the first conduction band CB1 at $k=A_0=F_0/\omega_0$, the second dashed blue line the band gap between CB3 and VB2 at $k=A_0$. The dashed red line indicates the predicted cutoff from the semi-classical electron-hole model  \cite{Semiclassical_many_elec}, and the dashed green lines the maximum energy differences between VB2 and CB1, CB2, CB3, respectively. Panels (b) and (c) show the total HHG spectra for two different field amplitudes. The meaning of the vertical lines is the same as in (a).}
\label{fig:high_occ_orbital_spec}
\end{figure}

\begin{figure}
\centering
\includegraphics[width=0.9\columnwidth]{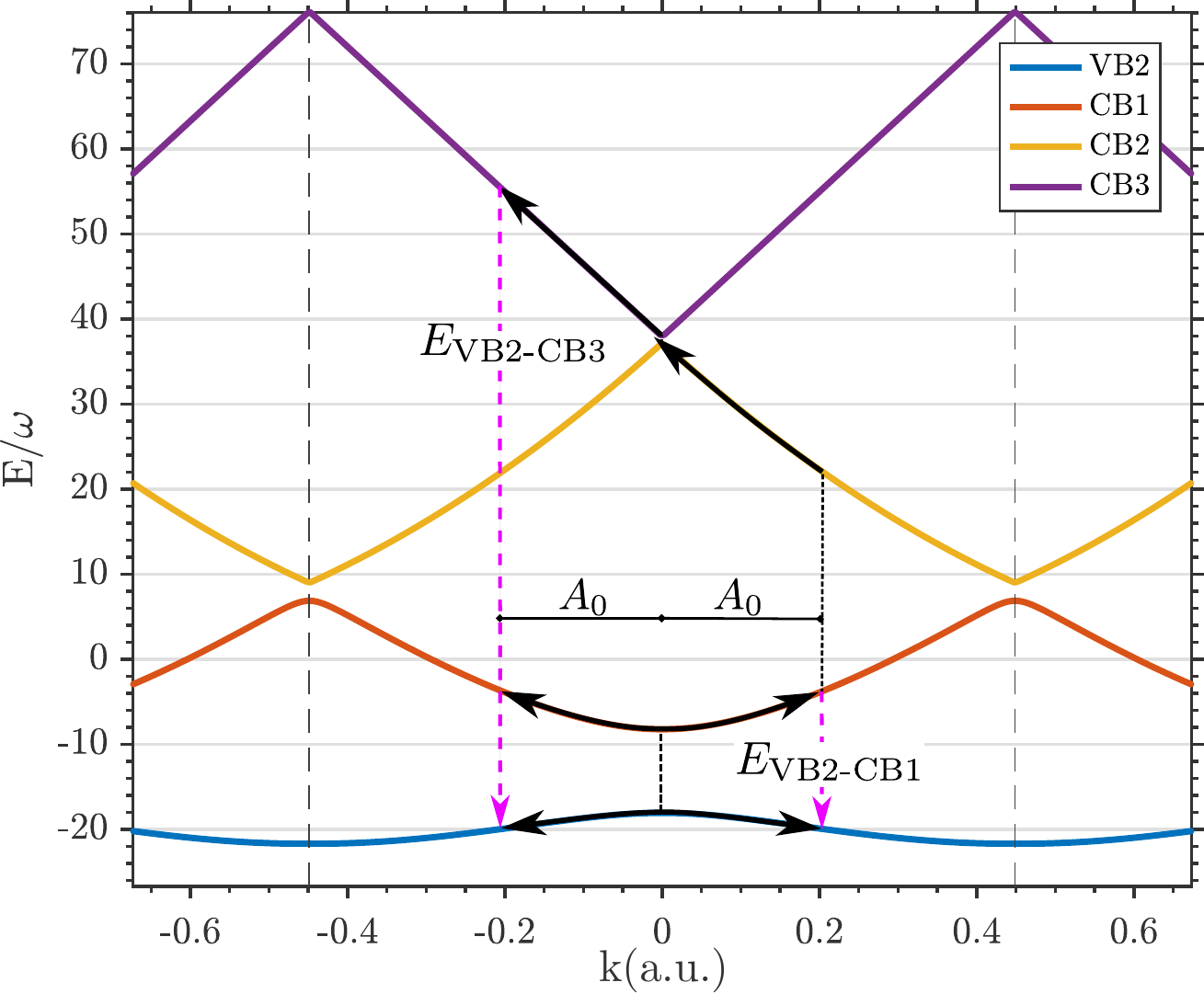}
\caption{The highest occupied valence band VB2 (blue) and the first three conduction bands CB1, CB2, CB3 (red, gold, purple) of system B with the energy plotted in units of the field frequency  $\omega_0$. The vertical gray, dashed lines indicate the Brillouin-zone boundary, the pink vertical dashed arrows indicate the maximum transition energies between VB2 and CB1, CB2, respectively. The black bold arrows indicate possible pathways of the highest occupied KS orbital (at $k_0=0$ in VB2) in the band structure. Vertical, black, dotted lines indicate tunneling transitions to next higher bands. }
\label{fig:bandstructure}
\end{figure}

Previous theoretical studies of HHG in a similar 1D system used an initial state made from a superposition of Bloch states around $k_0 = 0$ to predict HHG spectra \cite{gaarde_HHG}.
Figure~\ref{fig:high_occ_orbital_spec}a shows the HHG spectrum for the initially highest occupied KS orbital of system B, which is located in VB2 at $k_0=0$. Two cutoffs are indicated by dashed blue, vertical lines.  From the adiabatic theorem we expect that the $k_0=0$-state oscillates as $k(t) = A(t)$. This motion is indicated in Fig.~\ref{fig:bandstructure} by black bold arrows  in the valence band VB2 and the first conduction band CB1. If an electron makes a transition from VB2 to CB1 at $k=0$ and then continues to move in the field to $A_0$ (or $-A_0$) where it recombines with a hole in the initial band, the resulting maximum band gap is
\begin{align}
\omega_\mathrm{cutoff1} = E_{\textrm{CB1}}(A_0) - E_{\textrm{VB2}}(A_0).
\end{align}
The corresponding harmonic order $\omega_\mathrm{cutoff1}/\omega_0$  is indicated by the first dashed blue, vertical line from the left in Fig.~\ref{fig:high_occ_orbital_spec}a. When varying the laser intensity this point is found to follow the observed cutoff perfectly for the highest occupied orbital. The adiabatic theorem leads directly to the single-orbital cutoff from a semi-classical perspective of the initial state moving in the band structure with a certain maximum oscillation amplitude. The band gap energy gives the classically maximal obtainable energy, which therefore leads to a clear cutoff in the HHG spectrum.
A second cutoff is clearly identified in Fig.~\ref{fig:high_occ_orbital_spec}a and indicated by a second blue, vertical line. Varying the laser intensity, the cutoff was found to follow
\begin{align}
\omega_\mathrm{cutoff2} = E_{\textrm{CB3}}(A_0) - E_{\textrm{VB2}}(A_0). \label{eq:cutoff2}
\end{align}
With the vector potential amplitude $A_0=0.2$ used the $k_0=0$-state, according to the simple modeling, does not reach the Brillouin zone boundary at $k=\pi/7.0$ where the band gap to the next band CB2 is smallest. Hence less likely transitions from CB1 to CB2 at a larger energy gap need to take place. The question of how to calculate transition probabilities between bands has been addressed already in the second part of Keldysh's classic paper \cite{Keldysh1964} and, in view of the modern developments,  more recently in \cite{HawkinsIvanovPhysRevA.87.063842}. The  transition is indicated in Fig.~\ref{fig:bandstructure} by a dotted, vertical, black line. The transition from the second to the third conduction band requires only the transition through a small band gap of less than one photon energy.  HHG spectra from other individual orbitals starting from VB2 contain similar plateau structures but with less pronounced cutoffs as their initial momenta $k_0\neq 0$ enable them to explore more of the band structure. 

The full HHG spectrum of the entire system is presented in Fig.~\ref{fig:high_occ_orbital_spec}b. The lowest energy cutoff in the spectrum is located at the dashed red, vertical line. 
This cutoff is found to not follow a single orbital, e.g., the highest occupied KS orbital $k_0 = 0$. Instead, it moves linearly with the field strength.
The harmonic order for the dashed red line has been determined by finding the maximum recombination energy an electron-hole pair can have in a semi-classical model of interband HHG, as proposed by Vampa {\em et al.} \cite{Semiclassical_many_elec}. In this model, electron-hole pairs are created at the minimum band gap between VB2 and CB1 and then propagated in position space according to their respective dispersion relations. Later in the pulse the electron and hole can recollide at time $t_r$ and release the instantaneous energy difference between the bands at crystal momentum $k(t_r)$. 
We find that the total HHG spectra exhibit clear cutoffs at $\omega < E_\mathrm{CB1-VB2}$ in agreement with this semi-classical model (applied to all initial KS orbitals though)  for vector potential amplitudes $A_0 \lesssim 0.3$.  For higher amplitudes $A_0 > 0.3$ the cutoff shifts into the energy range of transitions from the second conduction band to the valence band, $E_\mathrm{CB1-VB2} < \omega < E_\mathrm{CB2-VB2}$ in accordance with an extended semi-classical hole-electron  (ESCHE) model  where electron transitions to higher conduction bands are taken into account whenever a minimum band gap is reached by a KS electron. 

In Fig.\ \ref{fig:high_occ_orbital_spec}b we also observe that harmonics below the 10th are suppressed when compared with the highest occupied orbital in Fig.\ \ref{fig:high_occ_orbital_spec}a. This is due to destructive interference as in Subsection~\ref{Blochoscillations} and in agreement with previous many-electron results from semi-conductor Bloch equations where a lack of peaks in the sub-threshold (i.e., below-bandgap) harmonics can also be observed \cite{Semiclassical_many_elec}.

A second cutoff---or at least a qualitative change in the total spectrum---can be observed at the second dashed blue line from the left in Fig.~\ref{fig:high_occ_orbital_spec}b which is located at $\omega_\mathrm{cutoff2}$ according \eqref{eq:cutoff2}, where the pronounced second cutoff for the initially highest occupied KS orbital occurs in  Fig.~\ref{fig:high_occ_orbital_spec}a. This high-energy feature is thus determined by a single orbital whereas the first, low-energy cutoff is not.

For  vector potential amplitudes  $A_0 > \pi/2a = 0.224$ (i.e., half the Brillouin zone boundary) the total spectrum changes character from three plateaus, as seen in Fig.~\ref{fig:high_occ_orbital_spec}b, to four plateaus in Fig.~\ref{fig:high_occ_orbital_spec}c.
The dashed red line again indicates the cutoff expected from the ESCHE model, which agrees well with the first cutoff from the TDDFT result. The dashed green lines indicate the maximum energy differences between VB2 and CB1, CB2, and CB3, respectively. They are already included in Figs.~\ref{fig:spectra_bands}a,b, showing that these maximum energy differences are not exhausted for small field strengths.
For $A_0=0.24$ instead, four plateaus are observed, which all have cutoffs located approximately at these maximum energy difference. This qualitative change in the spectrum suggests a change in the process leading to the second and third HHG plateau because transitions between bands at large energy gaps are not required anymore.

\begin{figure}
\medskip
\includegraphics[width=0.9\columnwidth]{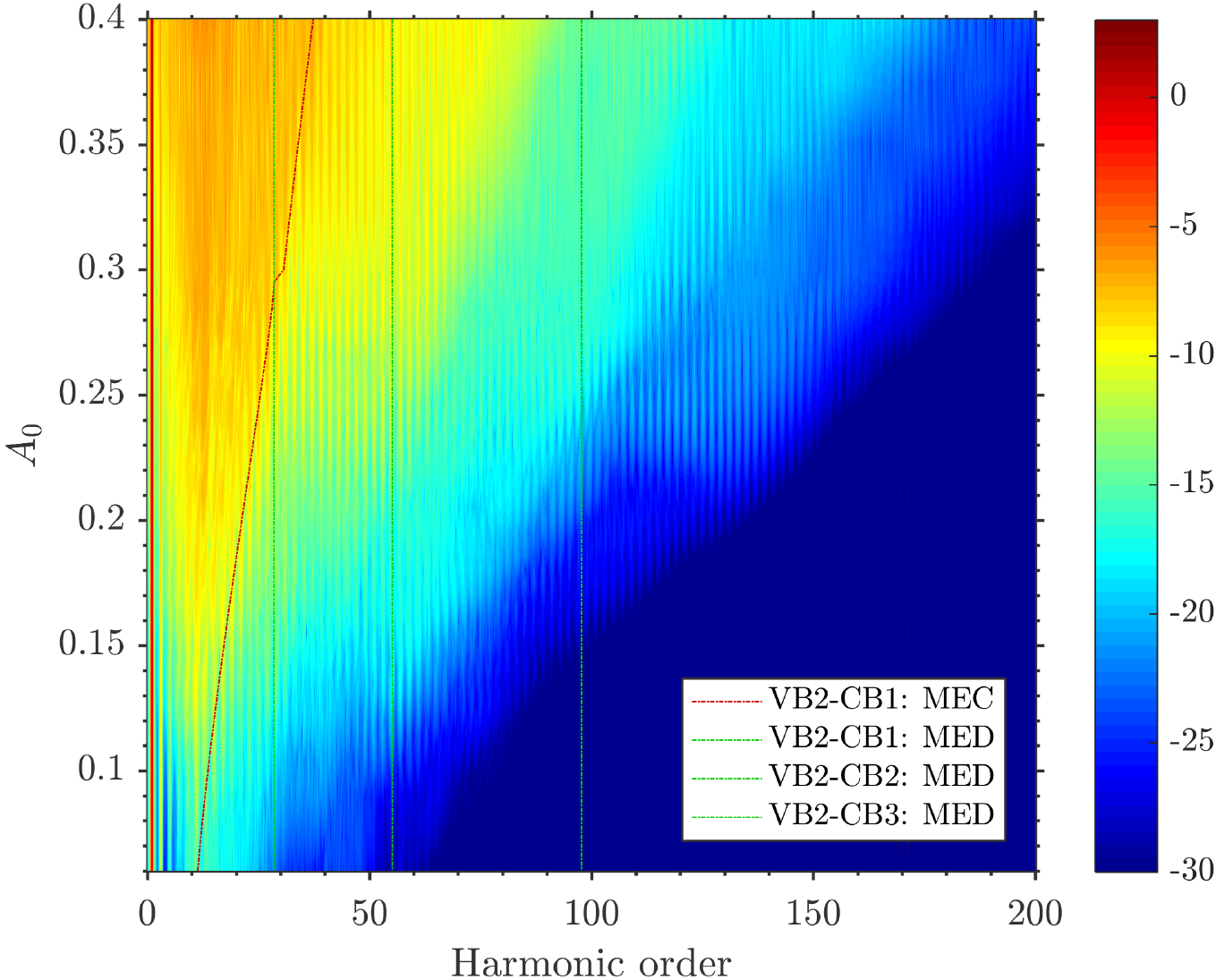}
\caption{HHG spectra for system B with frozen KS potential as a function of vector potential amplitude $A_0$. The red line indicates the prediction for the cutoff from the ESCHE model (labelled, MEC). The dashed green lines mark, from left to right, the maximum energy difference (MED) between VB2 and CB1, CB2, CB3.}
\label{fig:intensity_dependence}
\end{figure}

For  higher field intensities we confirm a cutoff scaling linear with $A_0$ found previously \cite{ishikawa_HHG,Semiclassical_many_elec}. In Fig.~\ref{fig:intensity_dependence} HHG spectra for system B are plotted as a function of the vector potential amplitude $A_0$ (the other laser parameters $\omega_0=0.023$ and $\ncyc=15$ in \eqref{eq:vpot} are kept the same). The overplotted red line indicates the first ESCHE cutoff as described in the discussion of Fig.~\ref{fig:high_occ_orbital_spec}. It jumps at $A_0 \simeq 0.3$ where the first cutoff moves into the second conduction band. The maximum energy differences between VB2 and CB1, CB2, CB3 are again highlighted by dashed green, vertical lines. 
A clear, linear cutoff scaling with $A_0$ can be inferred only for sufficiently high $A_0$ after the respective maximum band gaps have been exhausted.

\section{Summary}\label{sum}
We studied high-harmonic generation (HHG) in a simple 1D model system of a linear chain employing time-dependent density functional theory (TDDFT). In agreement with experiment and previous studies, multiple plateaus up to harmonic orders much higher than those obtained in gases are observed. These high harmonics are even observable when the vector potential amplitude is actually too small to drive the highest occupied Kohn-Sham (KS) orbital to the Brillouin zone boundary. This shows that over-simplified models where electrons only make transitions to the next band  at the smallest band gap are insufficient. 

The advantages of TDDFT over simpler models (where the time-dependent Schr\"odinger equation is solved for a single active electron moving in a given, periodic potential) are a full all-electron treatment, self-consistency, the incorporation of electron-electron interaction, and the ``automatic'' fulfillment of the Pauli exclusion principle. By comparing results from full TDDFT simulations with those for frozen Kohn-Sham (KS) potential we conclude that dynamic electron-electron interaction is of minor importance for HHG with the laser field strengths considered and up to harmonic orders with reasonable yields such that they could be of practical interest. Very important, instead, is the inclusion of all electrons in the valence band in the dipole or current from which the HHG spectra are calculated. HHG spectra from individual KS orbitals are in general very different from the full spectrum. An extreme case are Bloch oscillations in filled valence bands, which cancel almost completely.  We confirmed that the first cutoff in HHG spectra is indeed due to recombination of an electron from the first conduction band with the hole it left behind in the valence band. However, all KS electrons needed to be considered   to explain the dependence of the cut-off on the laser field strength. The higher plateau cutoffs are exhausted at photon energies corresponding to the maximum band gaps between higher-lying conduction bands and the valence band.  

The TDDFT model can obviously be extended in several directions, among them more complex unit cells, spin-polarized systems, classically mobile ions, 2D materials, and beyond the dipole approximation. The simplest next step is to consider edge effects, which will be the subject of a forthcoming paper.

\section*{Acknowledgments}
K.K.H. acknowledges support from the Villum-Kann Rasmussen (VKR) center of excellence QUSCOPE - Quantum Scale Optical Processes.

\bibliography{bibtex}

\end{document}